\documentclass[aps,twocolumn,nofootinbib,preprintnumbers,citeautoscript]{revtex4}
\usepackage{amsmath,amssymb,epsfig}
\usepackage{enumerate}
\usepackage{mathpazo}
\usepackage{braket}
\usepackage{amsmath}
\usepackage{color}
\usepackage{mathtools}
\DeclareMathOperator{\sinc}{sinc}
\usepackage[colorlinks=true,citecolor=blue,urlcolor=blue]{hyperref}
\newcommand{\ketbra}[2]{|#1\rangle \langle #2|}

\newcommand{\tr}{\operatorname{Tr}}
\newcommand{\be}{\begin{equation}}
\newcommand{\ee}{\end{equation}}
\newcommand{\ba}{\begin{eqnarray}}
\newcommand{\ea}{\end{eqnarray}}

\def\be{\begin{equation}}
\def\ee{\end{equation}}

\def\be{\begin{equation}}
\def\ee{\end{equation}}

\def\beq{\begin{eqnarray}}\def\eeq{\end{eqnarray}}

\usepackage{graphicx}

\begin {document}
\title{Entanglement certification and quantification in spatial-bin photonic qutrits}

\author{Debadrita Ghosh$^{1,\dagger}$, Thomas Jennewein$^{2}$, Urbasi Sinha$^{1}$}
\email{usinha@rri.res.in}
\affiliation{$^{1}$Light and Matter Physics, Raman Research Institute, Bengaluru-560080, India}
\altaffiliation{now at Laser-Laboratorium, Georg August Universit\" at G\" ottingen, Germany}
\affiliation{$^{2}$Institute for Quantum Computing, University of Waterloo, 200 University Avenue West, Waterloo, ON Canada}


\begin{abstract}

Higher dimensional quantum systems are an important avenue for new explorations in quantum computing as well as quantum communications. One of the ubiquitous resources in quantum technologies is entanglement. However, so far, entanglement has been certified in higher dimensional systems through suitable bounds on known entanglement measures. In this work, we have, for the first time, quantified the amount of entanglement in bi-partite pure qutrit states by analytically relating statistical correlation measures and known measures of entanglement, and have determined the amount of entanglement in our experimentally generated spatially correlated bi-partite qutrit system. We obtain the value of Negativity in our bi-partite qutrit to be 0.85 $\pm$ 0.03 and the Entanglement of Formation (EOF) to be 1.23 $\pm$ 0.01. In terms of quantifying the deviation from the maximally entangled state, the Negativity value demonstrates $\sim 15 \%$ deviation while the EOF value demonstrates $\sim 24\%$ deviation. This serves as the first experimental evidence of such non-equivalence of entanglement measures for higher dimensional systems.

\end{abstract}

\maketitle

\newpage

Entanglement \cite{Schrodinger} is one of the pivotal features of quantum mechanics that has deep-seated implications like nonlocality \cite{bell} and is a crucial resource for quantum communication and information processing tasks \cite{ben,deu,benn,JL03,ek,ABG+07,bruk,BCM+10,PAM+10,NPS14}. While the two-dimensional (qubit) case continues to be widely studied in the context of various applications of quantum entanglement, it has been gradually recognised that higher dimensional entangled states can provide significant advantages over standard two-qubit entangled states in a variety of cases, like, increasing the quantum communication channel capacity \cite{bennett,WDF+05}, enhancing the secret key rate and making the quantum key distribution protocols more robust in the presence of noise \cite{bech,CBK02,BCE+03,SV10} as well as enabling more robust tests of quantum nonlocality by reducing the critical detection efficiency required for this purpose \cite{ver}. 

Against the above backdrop, the enterprise of experimentally realising higher-dimensional entangled states, along with studies on the question of optimally certifying and quantifying higher-dimensional entanglement is of considerable importance. The usual method of characterising quantum states i.e. Quantum State Tomography (QST) or estimation of any entanglement measure  would require determination of an increasingly large number of independent parameters as the dimension of the system grows \cite{IOA17}. Therefore, formulating  experimentally efficient methods for the characterization of higher-dimensional entangled states based on limited number of measurements has become an active area of research \cite{TCA+14, gio,how,EKH17}. 

While many approaches for characterization of higher dimensional entanglement \cite{SSV13,SSV14,SHR17,SSC17,TDC+17,MGT+17,BVK+17,SH18,Roy05,DAC17,HMK+16,MBM15} have been studied (refer to \cite{PRAus} for an overview), only few schemes provide both necessary and sufficient certification together with quantification of high-dimensional entanglement. The quantification schemes that have been suggested so far seem to have focused essentially on providing bounds on entanglement measures. This gives rise to a need for exploring quantification of entanglement, by which, one essentially means  determining the actual value of an appropriate entanglement measure in terms of a limited number of experimentally measurable quantities using an analytically derived relation, for a given entangled state. Towards filling this gap, we apply a recently developed technique of experimentally generating spatially correlated bipartite photonic qutrits, whereby the quantumness of the correlated state generated is established through certification and quantification of entanglement. This is achieved for pure bipartite qutrits, by formulating and experimentally verifying analytical relations between statistical correlation measures and  known measures of entanglement namely Negativity and Entanglement of Formation. A salient feature of our work lies in our choice of correlation measures. On the one hand, we have employed commonly used correlation measures such as Mutual Predictability and Mutual Information; on the other hand, to the best of our knowledge we are the first team to use the Pearson Correlation Coefficient ($\mathcal{PCC}$) for higher dimensional entanglement characterization together with quantification. 

In this context, it is worth noting that common choices for photonic higher dimensional systems include those based on exploiting the Orbital Angular Momentum degree of freedom of a single photon \cite{PhysRevLett.116.073601,Palacios:11,Fickler13642,PhysRevA.76.042302,PhysRevA.69.023811}, spatial degree of freedom by placing apertures or spatial light modulators in the path of down-converted photons \cite{PhysRevA.78.012307,PhysRevA.79.043817,PhysRevA.80.062102,PhysRevLett.94.100501,PhysRevA.57.3123,PhysRevA.69.042305} as well as time-bin qudits \cite{PhysRevA.92.033802,PhysRevLett.93.180502,PhysRevA.66.062304}. 
Recently, our group has demonstrated a novel technique for spatial qutrit generation which is based on modulating the pump beam in spontaneous parametric down-conversion (SPDC) by appropriately placed triple slit apertures \cite{Ghosh:18}. This leads to direct generation of bipartite qutrits from the SPDC process which we call spatial-bin qutrits. This technique has been shown \cite{Ghosh:18} to be more efficient and robust, also leading to a more easily scalable architecture than what is achieved by the conventional method \cite{PhysRevA.86.012321} of placing slits in the path of down-converted photons. 
We used the pump-beam modulation technique for the first time in the qutrit domain yielding spatially correlated qutrits with a very high degree of correlation between measurements done in the image plane. However, no complementary basis (focal-plane) measurement was done to certify entanglement as has been argued to be necessary for such certification \cite{PhysRevA.57.3123}.\\
 In this work, what we have done has a three fold novelty. One is the {\it certification of entanglement} in pump beam modulated qutrits by appropriate measurements in complementary basis. The second is the main significance wherein we have shown in theory that we can relate statistical measures of correlations to known measures of entanglement i.e. Negativity($\mathcal{N}$) and Entanglement of Formation($\mathcal{EOF}$) through analytically derived monotonic relations and have applied these relations to {\it quantify} the produced entanglement in our experiment. Then the operationally relevant question arises as to how close is this prepared unknown state to the maximally entangled state? In this context, our work's final novelty lies in the comparison between the two inferred values of the two measures of entanglement i.e. $\mathcal{N}$ and $\mathcal{EOF}$ vis-a-vis their respective deviation from their defined values corresponding to the maximally entangled state. We find that these two measures of entanglement are non-equivalent in the sense that they yield different estimates of the deviation from the maximally entangled state. In a very recent study, the feature of non-equivalence has been comprehensively shown and analysed by us for two qubit pure states where the measures are seen to remain monotonically related \cite{2019arXiv190709268S}. However, in higher dimensional systems, studies have indicated that for certain classes of states, the different measures are not monotonically related to each other \cite{MG04, CS08, Ero15}. Ours is the first experimental demonstration of {\it non-equivalence} (in the sense defined above), between different measures of entanglement in higher dimensional bipartite pure states. An illustration of this feature along with attendant non monotonicity is provided in Appendix C in terms of theoretical estimates of the amounts of non equivalence for different choices of Schmidt coefficients for two qutrit pure states. This opens up interesting questions regarding the optimal choice of entanglement measure in different higher dimensional quantum information protocols. \\ 

Now, proceeding to the specifics of this paper, we first establish the relations between three statistical measures and two known measures of entanglement, followed by discussion of the experimental scheme we have used and the implications of the results obtained.

\section*{Deriving an analytic relation between $\mathcal{PCC}$ and  $\mathcal{N}$ for pure bipartite qutrit states}

$\mathcal{PCC}$ for any two random variables $A$ and $B$ is defined as

 \begin{equation} \label{PCCdef}
 \mathcal{C}_{AB}\equiv 
 \frac{\braket{AB}-\braket{A}\braket{B}}
 {\sqrt{\braket{A^2}-\braket{A}^2}\sqrt{\braket{B^2}-\braket{B}^2}},
 \end{equation}
whose values can lie between $-1$ and $1$, and  $\braket{\cdot}$ is an average value.

It is noteworthy that $\mathcal{PCC}$ has  so far been used in physics only in limited contexts \cite{BGP+17,BKO02} until recently when Maccone et al. \cite{MBM15} suggested its use for entanglement characterization.

Let us suppose two spatially separated parties, Alice and Bob, share a bipartite pure or mixed state in an arbitrary dimension; Alice performs two dichotomic measurements $A_1$ and $A_2$ and Bob performs two dichotomic measurements $B_1$ and $B_2$ on their respective subsystems. Then, Maccone et al. conjectured that the sum of two $\mathcal{PCC}$s being greater than $1$ for appropriately chosen mutually unbiased bases would certify entanglement of bipartite systems, i.e., for $A_1=B_1=\sum_j a_j \ketbra{a_j}{a_j}$ and $A_2=B_2=\sum_j b_j \ketbra{b_j}{b_j}$,
 \begin{equation}
|\mathcal{C}_{A_1B_1}|+|\mathcal{C}_{A_2B_2}|>1, \label{s2pcc}
 \end{equation}
would imply entanglement. Here, $\{\ket{a_j}\}$ is mutually unbiased to $\{\ket{b_j}\}$.  However, Maccone at al. justified this conjecture only by showing its applicability for bipartite qubits  and the validity of this conjecture has remained uninvestigated for dimensions $d>2$. In this work, we have justified the validity of this conjecture for pure bipartite qutrits by deriving an analytic relation between $\mathcal{PCC}$ and $\mathcal{N}$ and tested this relation by applying it to quantify the amount of entanglement in our experimentally generated pure bipartite qutrits. Please refer to \cite{PRAus} for a detailed study encompassing a wide range of mixed states for qutrits as well as qudits of higher dimension.

Consider a pure bipartite qutrit state written in Schmidt decomposition
\begin{equation}
\label{l1}
\ket{\psi}=c_{0}\ket{0}\ket{0}+c_{1}\ket{1}\ket{1}+c_{2}\ket{2}\ket{2}
\end{equation} 
where $\{\ket{0},\ket{1},\ket{2}\}$ are the computational bases. 

Let the basis $\{\ket{b_{j}}\}$ be the generalised $\hat{\sigma}_{x}$ basis \cite{SGB+06, SHB+12, PRAus}
\begin{align}
\label{l2}
\ket{b_{0}} & = \frac{1}{\sqrt{3}}[\ket{0}+\ket{1}+\ket{2}] \\
\label{l3}
\ket{b_{1}} & = \frac{1}{\sqrt{3}}[\ket{0}+\omega\ket{1}+\omega^{2}\ket{2}] \\
\label{l4}
\ket{b_{2}} & = \frac{1}{\sqrt{3}}[\ket{0}+\omega^{2}\ket{1}+\omega\ket{2}] 
\end{align}
where $\omega=e^{2i\pi/3}$ with $i=\sqrt{-1}$.

Consider a pair of observables $\hat{A_1}$ and $\hat{B_1}$ whose eigenstates are $\ket{b_{0}}, \ket{b_{1}}$ and $\ket{b_{2}}$ given by Eqs.(\ref{l2}),(\ref{l3}) and (\ref{l4}) respectively with the corresponding eigenvalues $b_{0}$, $b_{1}$ and $b_{2}$. Then we write 
\begin{align}
\label{e5}
\hat{A_1} & = \hat{B_1}=b_{0}\ket{b_{0}}\bra{b_{0}}+b_{1}\ket{b_{1}}\bra{b_{1}}+b_{2}\ket{b_{2}}\bra{b_{2}}
\end{align}
For the pair of observables given by Eq.(\ref{e5}) and the quantum state given by Eq.(\ref{l1}), one can use Eqs.(\ref{l2}),(\ref{l3}) and (\ref{l4}) to evaluate the quantity $\mathcal{C}_{A_1B_1}$ as given by 
\begin{equation}
\label{b1}
|C_{A_1B_1}|=\mathcal{N}
\end{equation}
where $\mathcal{N}$ corresponds to the Negativity measure of entanglement corresponding to the state given by Eq.(\ref{l1}) \cite{ETS15}. For  $A_2=B_2=\sum_j a_j \ketbra{a_j}{a_j}$ where the basis $\{\ket{a_j}\}$ is the computational basis, by computing the relevant single and joint expectation values for the pure two-qutrit states given by Eq.(\ref{l1}), one can obtain the $\mathcal{PCC}$ in this case to be given by 
\begin{equation}
\label{b2}
|C_{A_2B_2}|=1
\end{equation}
Thus, using Eqs.(\ref{b1}) and (\ref{b2}), we obtain,
\begin{equation}
\label{PCCsum}
|C_{A_1B_1}|+|C_{A_2B_2}| = 1 + \mathcal{N}\\
\end{equation}

Here one may just briefly remark that for deriving the value of $\mathcal{N}$ from the observed value of $\mathcal{PCC}$, it is empirically advantageous to consider single $\mathcal{PCC}$ because then the associated error range is less than when the sum of $\mathcal{PCC}$s is considered. Hence, from the point of view of such experimental consideration, in the following sections, we consider Mutual Predictability ($\mathcal{MP}$) and Mutual Information ($\mathcal{MI}$) as single quantities for analytically linking them to $\mathcal{N}$ and $\mathcal{EOF}$ respectively.

\section*{Relating Mutual Predictability with Negativity}

In showing the relation between $\mathcal{MP}$ and $\mathcal{N}$, we investigate the situation in which generalised $\hat{\sigma}_{x}$ observable is measured as one observable and its complex conjugate is measured as the other. We discuss the case when both observables are the same (see Appendix A).


Consider a pure bipartite qutrit state written in Schmidt decomposition as given in Eq.(\ref{l1}). 

For the complex conjugate basis of ${\ket{b_{j}}}$ and using $\omega^{*}=\omega^{2}$ and $(\omega^{2})^{*}=\omega$ we obtain $\ket{b_{0}}^{*} = \ket{b_{0}}$, $\ket{b_{1}}^{*} = \ket{b_{2}}$ and $\ket{b_{2}}^{*} = \ket{b_{1}}$. 

The above relations imply that one can obtain the probability of detecting the quantum state in the $\ket{b_{0}}^{*}$ or $\ket{b_{1}}^{*}$ or $\ket{b_{2}}^{*}$ state pertaining to the complex conjugate of generalised $\hat{\sigma}_{x}$ basis by using the generalised $\hat{\sigma}_{x}$ basis and obtaining the corresponding probability of detecting the quantum state in $\ket{b_{0}}$ or $\ket{b_{2}}$ or $\ket{b_{1}}$ state respectively.
Now, if we consider measuring generalised $\hat{\sigma}_{x}$ operator on one system and its complex conjugate on the other, then we can obtain the joint probabilities $P(b_{i},\bar{b}_{j})$ whence $P(b_{0},\bar{b}_{0}) = P(b_{1},\bar{b}_{1})=P(b_{2},\bar{b}_{2})$ (see Appendix B for details). The quantities $P(b_{1},\bar{b}_{1})$, $P(b_{2},\bar{b}_{2})$ in this case are same as the quantities $P(b_{1},b_{2})$ and $P(b_{2},b_{1})$ as measured by using generalised $\hat{\sigma}_{x}$ operator on both systems. 
Then, one can obtain \cite{SHB+12} the $\mathcal{MP}$ as
\begin{equation}
\label{l5}
\mathcal{C}  = \sum_{i}P(b_{i},\bar{b}_{i})
= \frac{1}{3}(1+2\mathcal{N})
\end{equation}
where $\mathcal{N}$ is the negativity of the bipartite qutrit state \cite{ETS15}.



\section*{Relating Mutual Information with Entanglement of Formation}

Let the common basis  of the  pair of observables $A$ and $B$ pertaining to Alice and Bob 
be the computational basis. For this choice of measurements, let the joint probabilities be $p(ab|AB)$ and the marginal probabilities be $p(a|A)$ and $p(b|B)$  for the pure two-qudit state
\begin{equation}
\ket{\psi_d}=\sum^{d-1}_{i=0}c_i \ket{ii}, \label{pqud}
\end{equation}
where $0 \le c_i \le 1$ and $\sum_ic^2_i=1$,
are given by $p(ab|AB)= c_i^2$ for $a=b=i$ and $0$ otherwise while $p(a|A)=p(b|A)=c_i^2$ for $a=b=i$.
Substituting the above joint probabilities and marginal probabilities in the expression for $\mathcal{MI}$ \cite{MBM15} given by 
\be
I_{AB}=\sum^{d-1}_{a,b=0}p(ab|AB) \log_2\frac{p(ab|AB)}{p(a|A)p(b|B)},
\ee
we obtain
\begin{equation}
 I_{AB}=-\sum^{d-1}_{i=0}c^2_i\log_2c^2_i \label{MI}
\end{equation}

Now, note that $\mathcal{EOF}$, for bipartite pure states $\ket{\psi}_{AB}$
is equal to the von Neumann entropy of either of the reduced 
density matrices, i.e.,
 $\mathcal{E}(\ket{\psi}_{AB})=S(\rho_A)=S(\rho_B)$, here 
 $S(\rho)=-\tr\rho \log_2\rho$. For the general pure two-qudit state as 
 given by Eq.(\ref{pqud}), $\mathcal{EOF}$ is 
given by the following expression:
\be \label{eofPtQd}
\mathcal{E}(\ket{\psi_d})=-\sum_i c_i^2 \log_2 c^2_i.
\ee
since $S(\rho_A)=S(\rho_B)=-\sum_i c_i^2 \log_2c^2_i$.
From Eqs.(\ref{MI}) and (\ref{eofPtQd}), it then follows that $\mathcal{MI}$ pertaining to the computational basis on both
sides equals the $\mathcal{EOF}$ for any pure bipartite qudit state.

\section*{Experimental scheme}
 \begin{figure}
\centering
\includegraphics[scale=0.16]{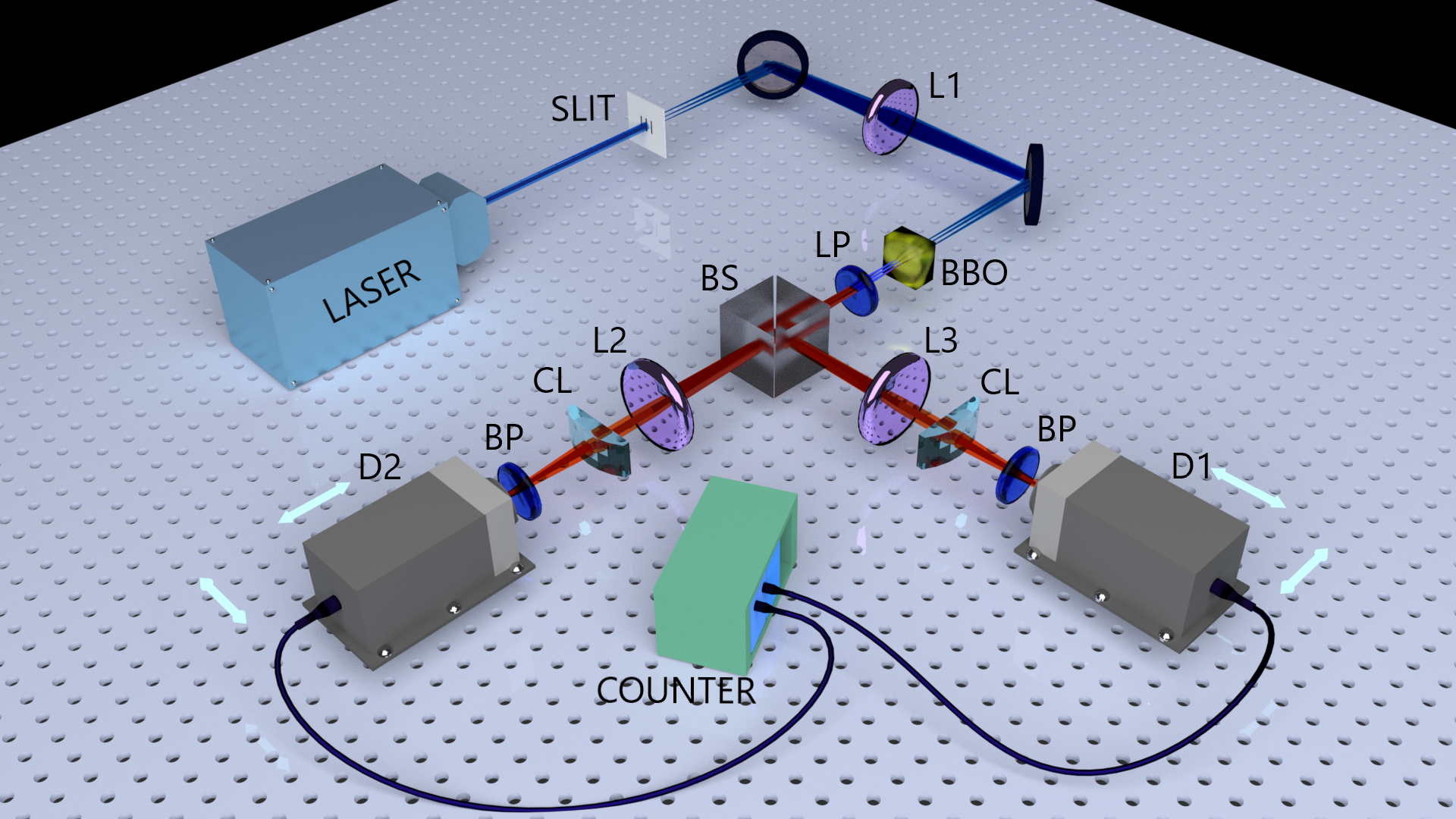}
\caption{Schematic of the experimental set-up. L1, L2, L3: Plano convex lenses, BBO: Nonlinear crystal for SPDC, LP: Long-pass filter, BS: 50-50 Beamsplitter, BP: Band-pass filter, CL: Cylindrical Lens, D1, D2: Single photon detectors. }
\label{}
\end{figure}

\begin{figure}
\centering
\includegraphics[scale=0.300]{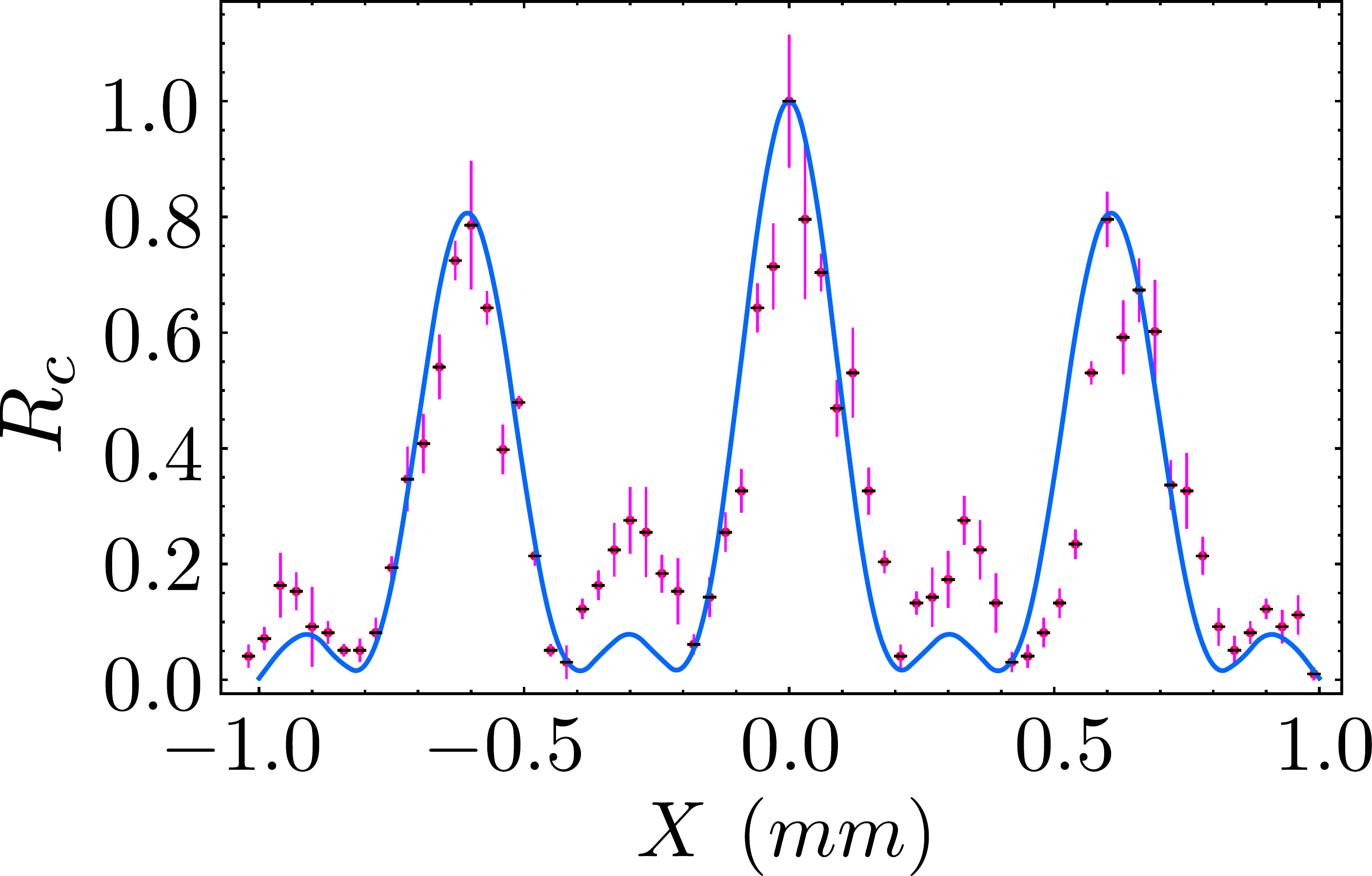}
\caption{Normalized coincidence count (Rc) vs detector position (X) in focal plane. Blue line indicates the theoretical prediction whereas the red circle indicates the experimental result. The normalised coincidence plot exhibiting interference is a certification of entanglement.}
\label{}
\end{figure}

\begin{figure*}
	\includegraphics[width=1\textwidth]{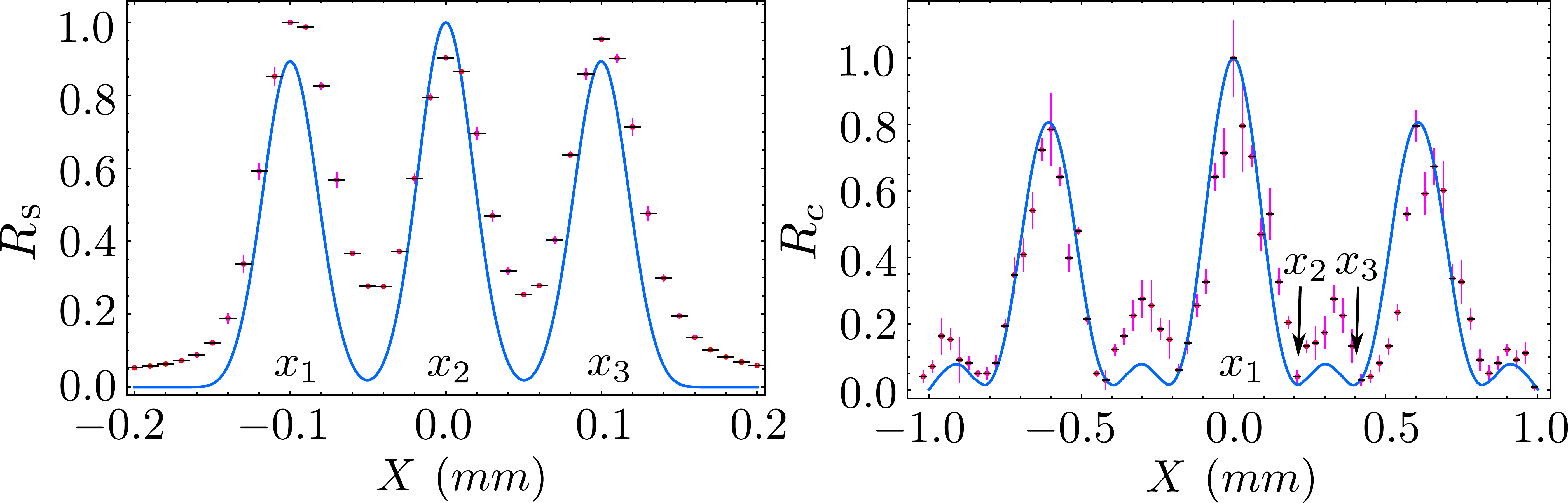}
	\caption{Normalized singles count (Rs) vs detector position (X) measured in image plane (left).  Normalized coincidence count (Rc) vs detector position (X) measured in focal plane (right). Here, the signal arm detector is fixed at the centre of the singles profile (called $y_1$ in text) while the idler arm is scanned. $x_1, x_2$ and $x_3$ in the plots represent the positions derived from the generalized $\sigma_z$ and  $\sigma_x$ like operators respectively.}  
\label{fig:combined}
\end{figure*}

A Type-1 BBO source is used to generate spatially correlated qutrit pairs using our previously developed pump beam modulation technique \cite{Ghosh:18}. Appendix D has details on the source. We explore the system (schematic in Fig. 1) at two different positions, i.e. focal (f) and image plane (2f) of the lenses L2 and L3. When the detectors are placed at the focus of lenses L2 and L3, the lenses transfer the triple-slit interference in the correlation of the signal and idler photons to the detector plane. In addition, there are two cylindrical lenses of focal lengths 50 mm and 60 mm in the transmitted (let's call signal arm) and reflected (let's call idler arm) arms of BS respectively which focus the single photons along a line at the detector plane. The individual measured singles spatial profiles of signal and idler photons have a flat top Gaussian structure. \\
We fix one detector at the centre of an individual profile and move the other detector to measure the coincidence. The moving detector is scanned over 4 mm in 30  $\mu m$ step size. At each position, three counts are recorded, with an accumulation time of 90 sec each. Fig.2 shows the coincidence profile measured in the interference plane. The blue line represents the theoretical prediction while the red circles are the measured values. We include error bars in terms of position and number uncertainty. The error in the position is limited by the step size of the actuator which moves the detector. The chosen step sizes of the actuator to measure the profiles are 10  $\mu m$ and 30  $\mu m$ for image plane and focal plane respectively. Experimental and theoretically generated Rc  have been appropriately normalised by their respective maxima. The focal plane measurement shown in Fig.2 has an important significance. As discussed in \cite{PhysRevA.57.3123}, when cross correlation measurements (also called coincidence measurements) as a function of detector position in the focal plane for both the signal and idler photons exhibit interference, this implies certification of entanglement.\\
The image plane corresponds to the generalised $\sigma_z$ like operator with eigenstates comprising of the computational basis states (discussed in theory section above). The three slit peak positions represent the three eigen states with eigen values 0, 1 and -1 respectively. In order to calculate the $\mathcal{PCC}$ for generalised $\sigma_z$ like operators applied to both the signal and the idler photons, we need to measure the corresponding joint probabilities. We fix one detector at the three peak positions, one position at a time, of its singles spatial profile (let’s say signal arm) and measure the coincidence counts when the other detector is at the peak positions of its singles profile (idler arm). Thus, by measuring the peak to peak coincidence counts we construct a 3 x 3 matrix with 9 components. The maximum coincidence counts are the diagonal elements of the matrix. We measure 5 such sets of matrices and find the average $\mathcal{PCC}$ to be 0.904(2). Table I (left) shows a representative correlation matrix.\\
The focal plane corresponds to the generalised $\sigma_x$ like operator with eigenstates given by Eqs.(\ref{l2}), (\ref{l3}) and (\ref{l4}). These correspond to three unique positions in the measured cross correlation profile (see Appendix E). In order to measure the $\mathcal{PCC}$ for generalised $\sigma_x$ like operators applied to both the signal and the idler photons, we need to measure the corresponding joint probabilities like in the case of $\sigma_z$ like operator. To construct such a joint probability matrix, we measure the singles profile for the signal arm and by fitting it with a flat-top Gaussian module function, we find out the centre position naming it $y_1$ . Similarly, we find the centre position for the idler singles profile and name it $x_1$ . Next, we fix the signal arm at $y_1$ and scan the idler arm to measure the coincidence profile. From this, we extract the positions $x_1$ , $x_2$ and $x_3$ corresponding to the eigenstates of the generalised $\sigma_x$ like operator. Next, we fix the idler arm at $x_1$ , $x_2$ and $x_3$ respectively and for each fixed position we scan the signal arm to measure the coincidence profiles. From each of the coincidence profiles, we extract the coincidence counts at the positions $y_1$ , $y_2$ and $y_3$ which are the derived positions for generalized $\sigma_x$ like operator for signal arm. \\
Fig. 3 represents the detector positions corresponding to generalised $\sigma_z$ and $\sigma_x$ eigen states respectively. Thus we construct a 3x3 correlation matrix. We measure 5 such sets of matrices and find the average $\mathcal{PCC}$ to be 0.848(2). Table I (right) shows a representative matrix. We work with operators defined by assigning eigen values 0, 1 and -1 respectively to the eigen states of the $\sigma_x$ like operator.\\
Then by invoking Eq.(\ref{b1}) derived earlier, we derive $\mathcal{N}$ as a measure of entanglement to be 0.848 $\pm$ 0.027.\\ 
Next we calculate $\mathcal{N}$ from measured $\mathcal{MP}$. As shown before, when generalised $\sigma_x$ basis is measured on one side (say the signal arm) and its complex conjugate basis is measured on the other side (idler arm), then the quantities $P(b_{1},\bar{b}_{1})$, $P(b_{2},\bar{b}_{2})$ in this case are same as the quantities $P(b_{1},b_{2})$ and $P(b_{2},b_{1})$ respectively, as measured by using generalised $\hat{\sigma}_{x}$ basis on both sides. Thus summing over the $x_1 - y_1$, $x_2 - y_3$ and $x_3 - y_2$ elements of the correlation matrix as represented in Table I, one can derive $\mathcal{MP}$.  The $\mathcal{MP}$ so derived comes out to be 0.899 $\pm$ 0.013 with the $\mathcal{N}$ derived using Eq.(\ref{l5}) equal to 0.849 $\pm$ 0.020.\\
Using Eq.(\ref{MI}) we calculate the $\mathcal{MI}$ where $c_i$ are the normalised coincidence counts when one detector is fixed at $x_1$ (then $x_2$ and $x_3$) and the other detector moves from $y_1$ to $y_3$ respectively. $\mathcal{MI}$ is equal to the $\mathcal{EOF}$ in case of computational basis as shown earlier.  Our calculated $\mathcal{EOF}$ is 1.23 $\pm$ 0.01.\\

It is thus clear from Table II that $\mathcal{N}$ as a measure of entanglement derived from two independent statistical correlation measures is the same within error bound. An interesting point emerges here. All the derivations shown in this manuscript as well as experimentally derived quantities assume the initial bi-partite qutrit state to be a pure state. The value of $\mathcal{N}$ being the same within error bound serves as a consistency check for this assumption. Moreover, we know that the concept of coherence is intimately connected with mixedness of the state \cite{PhysRevA.92.022316} and the measure of coherence is the Visibility of the interference \cite{PhysRevLett.85.2845}. Higher degree of coherence implies higher state Purity. In our experiment, the interference in the coincidence plane has a Visibility of $\sim 94 \%$ which indicates a high state Purity. The question now arises: what is the percentage deviation from the maximally entangled state as quantified by two different entanglement measures i.e.  $\mathcal{N}$ and $\mathcal{EOF}$? Noting that the maximum values of $\mathcal{N}$ and $\mathcal{EOF}$ are 1 and $\log_2 3=1.585$ respectively, it is found that while the $\mathcal{N}$  value quantifies a $\sim 15 \%$ deviation from the maximally entangled state, the deviation captured by the $\mathcal{EOF}$ is around $\sim 24 \%$. Thus, we conclusively demonstrate in an experimental scenario, a significant amount of non-equivalence between different measures of entanglement in higher dimensions.\\
In conclusion, we have certified entanglement in our pump beam modulated spatially correlated bipartite qutrits. We have then developed a novel method in theory and applied the same to quantify the amount of entanglement in our bipartite qutrits using $\mathcal{N}$ and $\mathcal{EOF}$ as measures of entanglement. Our method of entanglement characterisation is applicable to unknown quantum states involving limited number of measurements as opposed to extracting such information from a complete QST. This has led to a curious observation that the two measures of entanglement differ strikingly in quantifying the deviation of entanglement in the prepared unknown state from that in the maximally entangled state. This, in conjunction with the feature of non monotonicity illustrated by us (see Appendix C) calls for a deeper understanding for the underlying reasons and reexamining the efficacy of these measures for quantifying the resource for higher dimensional quantum information processing protocols. 

\begin{table}
 \begin{tabular}{||c| c| c| c||} 
 \hline
 & $x_1$ & $x_2$ & $x_3$\\ [0.5ex] 
 \hline\hline
 $y_1$ & 0.281 & 0.024 & 0.003 \\ 
 \hline
 $y_2$ & 0.006 & 0.287 & 0.014 \\
 \hline
 $y_3$ & 0.002 & 0.006 & 0.376 \\
 \hline
  
\hline
\end{tabular}
\quad
 \begin{tabular}{||c| c| c| c||} 
 \hline
 & $x_1$ & $x_2$ & $x_3$\\ [0.5ex] 
 \hline\hline
 $y_1$ & 0.344 & 0.017 & 0.017 \\ 
 \hline
 $y_2$ & 0.008 & 0.017 & 0.260 \\
 \hline
 $y_3$ & 0.017 & 0.302 & 0.017 \\
 \hline
\end{tabular}
\caption{3x3 correlation matrix corresponding to generalized $\sigma_z$ like operator measured in image plane (left) and $\sigma_x$ like operator measured in focal plane (right).}
\end{table}
\begin{table}
\begin{center}
 \begin{tabular}{||c| c| c| c||} 
 \hline
  $\mathcal{N}$ from $\mathcal{PCC}$ & $\mathcal{N}$ from $\mathcal{MP}$ & $\mathcal{E}$ from $\mathcal{MI}$ \\ [0.5ex] 
 \hline\hline
  0.848$\pm 0.027$ & 0.849$\pm 0.020$ & 1.233$\pm 0.012$ \\
 \hline
\end{tabular}
\caption{Values of different measures of entanglement.}
\end{center}
\end{table}


\begin{acknowledgments}

The authors thank C. Jebarathinam and D.Home for useful discussions on theoretical concepts. US thanks P. Kolenderski, A.K.Pati and A.Sinha for useful discussions. Thanks are due to S. Bhar for initial discussions.

\end{acknowledgments}

\bibliography{new}

\section*{Appendices}


\subsection{Mutual Predictability when generalized $\sigma_{x}$ observable is measured on both sides}
Consider a pure bipartite qutrit state written in Schmidt decomposition
\begin{equation}
\label{e1}
\ket{\psi}=c_{0}\ket{0}\ket{0}+c_{1}\ket{1}\ket{1}+c_{2}\ket{2}\ket{2}
\end{equation} 
where $\{\ket{0},\ket{1},\ket{2}\}$ are the computational bases. 

Let the basis $\{\ket{b_{j}}\}$ be the generalized $\hat{\sigma}_{x}$ basis \cite{SGB+06, SHB+12, PRAus}
\begin{align}
\label{e2}
\ket{b_{0}} & = \frac{1}{\sqrt{3}}[\ket{0}+\ket{1}+\ket{2}] \\
\label{e3}
\ket{b_{1}} & = \frac{1}{\sqrt{3}}[\ket{0}+\omega\ket{1}+\omega^{2}\ket{2}] \\
\label{e4}
\ket{b_{2}} & = \frac{1}{\sqrt{3}}[\ket{0}+\omega^{2}\ket{1}+\omega\ket{2}] 
\end{align}
where $\omega=e^{2i\pi/3}$ with $i=\sqrt{-1}$.
\begin{equation}
\label{e7.1}
P(b_{i},b_{j})=|\bra{\psi}\ket{b_{i}}\bra{b_{i}}\otimes\ket{b_{j}}\bra{b_{j}}\ket{\psi}|^{2}
\end{equation}
Using Eq.(\ref{e7.1}) we obtain the following joint probabilities

\begin{align}
\label{s1}
P(b_{0},b_{0}) & = P(b_{1},b_{2})=P(b_{2},b_{1})\\&=\frac{1}{9}(1+2c_{0}c_{1}+2c_{2}c_{1}+2c_{0}c_{2})\\
\label{s2}
P(b_{1},b_{1}) & = P(b_{2},b_{2})=P(b_{0},b_{2})\\&=P(b_{2},b_{0})=P(b_{0},b_{1})=P(b_{1},b_{0})\\&=\frac{1}{9}(1-c_{0}c_{1}-c_{2}c_{1}-c_{0}c_{2})
\end{align}
It can be checked from Eqs.(\ref{s1}) and (\ref{s2}) that $\sum_{i,j}P(b_{i},b_{j})=1$.\\

Using the relevant joint probabilities from Eqs.(\ref{s1}) and (\ref{s2}) one can then obtain Mutual Predictability \cite{SHB+12} as 
\begin{align}
\label{a1}
\mathcal{C} & = \sum_{i}P(b_{i},b_{i})\\
\label{a2}
& = \frac{1}{3}
\end{align}

\subsection{Mutual Predictability when generalized $\sigma_{x}$ observable is measured on one side and its complex conjugate is measured on the other side}

Consider a pure bipartite qutrit state written in Schmidt decomposition as given in Eq.(\ref{e1}). \\
One can construct the complex conjugate bases of $\{\ket{b_{j}}\}$ as follows
\begin{align}
\label{e12}
\ket{b_{0}}^{*} & = \frac{1}{\sqrt{3}}[\ket{0}+\ket{1}+\ket{2}] \\
\label{e13}
\ket{b_{1}}^{*} & = \frac{1}{\sqrt{3}}[\ket{0}+\omega^{*}\ket{1}+(\omega^{2})^{*}\ket{2}] \\
\label{e14}
\ket{b_{2}}^{*} & = \frac{1}{\sqrt{3}}[\ket{0}+(\omega^{2})^{*}\ket{1}+\omega^{*}\ket{2}] 
\end{align}

where $\omega^{*}$ is the complex conjugate of $\omega$. Using $\omega^{*}=\omega^{2}$ and $(\omega^{2})^{*}=\omega$ we obtain
\begin{align}
\label{s12}
\ket{b_{0}}^{*} & = \ket{b_{0}} \\
\label{s13}
\ket{b_{1}}^{*} & = \ket{b_{2}} \\
\label{s14}
\ket{b_{2}}^{*} & = \ket{b_{1}} 
\end{align}

Eqs.(\ref{s12})-(\ref{s14}) imply that one can obtain the probability of detecting the quantum state in the $\ket{b_{0}}^{*}$ or $\ket{b_{1}}^{*}$ or $\ket{b_{2}}^{*}$ state pertaining to the complex conjugate of generalized $\hat{\sigma}_{x}$ basis by using the generalized $\hat{\sigma}_{x}$ basis and obtain the corresponding probability of detecting the quantum state in $\ket{b_{0}}$ or $\ket{b_{2}}$ or $\ket{b_{1}}$ state respectively.\\

Now, if we consider measuring generalized $\hat{\sigma}_{x}$ on one side and its complex conjugate on the other side, then we can obtain the joint probabilities $P(b_{i},\bar{b}_{j})$ as
\begin{equation}
\label{s7.1}
P(b_{i},\bar{b}_{j})=|\bra{\psi}\ket{b_{i}}\bra{b_{i}}\otimes\ket{b_{j}^{*}}\bra{b_{j}^{*}}\ket{\psi}|^{2}
\end{equation}
where $\bar{b}_{j}$ denotes the eigenvalue corresponding to the complex conjugate of the $\ket{b_{j}}$, whence we obtain
\begin{align}
\label{ss1}
P(b_{0},\bar{b}_{0}) & = P(b_{1},\bar{b}_{1})=P(b_{2},\bar{b}_{2})=\frac{1}{9}(1+2c_{0}c_{1}+2c_{2}c_{1}+2c_{0}c_{2})
\end{align}
As discussed in the preceding paragraph, the quantities $P(b_{1},\bar{b}_{1})$, $P(b_{2},\bar{b}_{2})$ in this case are same as the quantities $P(b_{1},b_{2})$ and $P(b_{2},b_{1})$ given by Eq. (\ref{s1}) respectively, as measured by using generalized $\hat{\sigma}_{x}$ basis on both sides.\\

Now, using Eq. (\ref{ss1}), one can obtain \cite{SHB+12} the Mutual Predictability as 
\begin{align}
\label{a11}
\mathcal{C} & = \sum_{i}P(b_{i},\bar{b}_{i})\\
\label{a12}
& = \frac{1}{3}(1+2c_{0}c_{1}+2c_{2}c_{1}+2c_{0}c_{2})\\
\label{a13}
& = \frac{1}{3}(1+2\mathcal{N})
\end{align}
where $\mathcal{N}$ is the negativity of the bipartite qutrit state \cite{ETS15}.

\subsection{Analytical study of Relationship between $\mathcal{N}$ and Entanglement of Formation($\mathcal{E}$) for two qutrit pure states}

Consider a two qutrit pure state with Schmidt coefficients $c_{0}$ , $c_{1}$ \& $c_{2}$ 
\begin{align}
    \ket{\Phi} = c_{0}\ket{0}\ket{0} + c_{1}\ket{1}\ket{1} + c_{2}\ket{2}\ket{2} 
\end{align}
where $0 < c_{0}$,$c_{1},c_{2} < 1$ and $c_{0}^2$ + $c_{1}^2$  + $c_{2}^2$ = 1

\subsection*{Deviations from the Maximally Entangled State}
Two parameters are defined to measure the percentage deviations of measures from  the values corresponding to Maximally Entangled State \cite{2019arXiv190709268S}

\begin{align}
Q_{E} = ((log_{2}(3) - E)/log_{2}(3))\times 100 
\end{align}
\begin{align}
 Q_{N} = (1 - N))\times 100    
\end{align}

To see to what extent these two parameters differ with each other, the following quantity is an appropriate measure \cite{2019arXiv190709268S}
\begin{align}
 \Delta Q_{NE} = |Q_{E}- Q_{N}|
 \end{align}
    
\subsection*{Study of Monotonicity}
 Rate of change of $\mathcal{E}$ w.r.t $c_{0}$
\begin{equation}
\dfrac{\mathrm{d}\mathcal{E}}{\mathrm{d}c_{0}}= (2/ln(2))c_{0}log_{2}(1-(c_{0}^2+c_{1}^2))/c_{0}^2) \label{m1}
\end{equation}
Similarly, rate of change of $\mathcal{E}$ w.r.t $c_{1}$
\begin{equation}
   \dfrac{\mathrm{d}\mathcal{E}}{\mathrm{d}c_{1}} = (2/ln(2))c_{1}log_{2}(1-(c_{0}^2+c_{1}^2))/c_{1}^2) \label{m2}
\end{equation}
 Rate of change of $\mathcal{N}$ w.r.t $c_{0}$
\begin{equation}
   \dfrac{\mathrm{d}\mathcal{N}}{\mathrm{d}c_{0}} = c_{1} + (1-c_{0}c_{1}-c_{1}^2-2c_{0}^2)/\sqrt{1-(c_{0}^2+c_{1}^2)} \label{m3}
\end{equation}
Similarly, rate of change of $\mathcal{N}$ w.r.t $c_{1}$
\begin{equation}
   \dfrac{\mathrm{d}\mathcal{N}}{\mathrm{d}c_{1}} = c_{0} + (1-c_{0}c_{1}-c_{0}^2-2c_{1}^2)/\sqrt{1-(c_{0}^2+c_{1}^2)} \label{m4}
\end{equation}
\textbf{Observations}: 
\begin{itemize}
    \item From Eqs. (\ref{m1}),(\ref{m2}),(\ref{m3}),(\ref{m4}) it can be seen that for a given $c_{0}$ ($c_{1}$), $\mathcal{E}$ and $\mathcal{N}$ grow with $c_{0}$ ($c_{1}$), reach a certain value and then start decreasing w.r.t $c_{0}$ ( $c_{1}$). All the above four Eqs.(\ref{m1}),(\ref{m2}),(\ref{m3}),(\ref{m4}) vanish when $c_{0}$ = $c_{1}$ = $1/\sqrt{3}$
    
    \item As in the case of two qubit pure state \cite{2019arXiv190709268S}, one cannot say that $\dfrac{\mathrm{d}\mathcal{E}}{\mathrm{d}\mathcal{N}}$ is always greater than zero except when $c_{0}$ \& $c_{1}$ = $1/\sqrt{3}$. This explains the presence of non-monotonic nature between these two parameters. For example, consider a pair of two qutrit pure states with Schmidt coefficients $c_{0}$ = 0.4 , $c_{1}$ = 0.9 \& $c_{0}$ = 0.5 , $c_{1}$ = 0.1. Former state has, 
    $$ E_{1} = 0.8879\  \&\  N_{1} = 0.5661 $$
    whereas the latter state has 
    $$ E_{2} = 0.8210\  \&\  N_{2} = 0.5852 $$
    Here $E_{1} > E_{2}$ but $N_{1} < N_{2}$, showing $\mathcal{E}$ and $\mathcal{N}$ are not monotonic w.r.t each other.

\end{itemize}

\subsection*{Study of Deviations from the Maximally Entangled State}
\begin{itemize}
    \item Different values of $\Delta Q$ for different values of the Schmidt coefficients have been tabulated in Table III. 
    \begin{table}[h]
    \begin{center}
    \caption{Differences in the \% deviations from the value corresponding to the maximally entangled state}
    \label{tab:table1}
    \begin{tabular}{l|c|c|c|c|c|r} 
      ${c_{0}}$ & ${c_{1}}$ & \textbf{E} & \textbf{N} & ${Q_{E}}$ & ${Q_{N}}$  & {$\Delta Q_{NE}$}\\
      
      \hline
      
      0.1 & 0.1 & 0.1614 & 0.2080 & 89.8142 & 79.2010 & 10.6132 \\
      0.3 & 0.8 & 1.2347 & 0.8116 & 22.0964 & 18.8423 & 3.2540 \\
      0.5774 & 0.5774 & 1.5850 & 1 & 0 & 0 & 0 \\
      0.6 & 0.6 & 1.5755  & 0.9950  & 0.6001 & 0.5020 & 0.0982\\
      0.9 & 0.3 & 0.8911 & 0.6495 & 43.7784 & 35.0527 & 8.7257\\

    \end{tabular}
     \end{center}
    \end{table}
    
     \textbf{Observations} :
     \item Given a non-maximally entangled two qutrit pure state, one cannot comment as in the case of two qubit case \cite{2019arXiv190709268S} that one entanglement measure is always greater than other entanglement measure for any value of state parameter. 
     
     \item $\Delta Q_{NE}$ takes a maximum value of 12.148\% when $c_{0} = 0.1712$ \&  $c_{1} = 0.1712$ 
    \item Thus, both as absolute and relative entanglement measures, Negativity and Entanglement of Formation do not give equivalent results. 

\end{itemize}

\subsection{ Details on the photon source and pump beam modulation technique}

A Type-1, non-linear crystal (BBO) with a dimension of 5 mm x 5 mm x 10 mm generates parametrically down-converted degenerate photons at 810 nm wavelength with collinear phase matching condition. A diode laser at 405 nm with 100 mw power pumps the crystal. The transverse spatial profile of the pump beam at the crystal is prepared by transferring the laser beam
through a three-slit aperture with 30 $\mu m$ slit width and 100  $\mu m$ inter-slit distance and imaging it at the crystal. A plano-convex lens (L1) of focal length 150 mm is placed such that it forms the image of the three-slit at the crystal. A 50-50 beam-splitter (BS) placed after the crystal splits the two down-converted photons in the transmitted and reflected ports of the BS. A band-pass filter centred at 810 nm with a FWHM of 10 nm passes the down-converted photons and a long-pass filter with cut-off wavelength 715 nm blocks the residual pump. In each arm of the BS, a plano-convex lens (L2 and L3 respectively) of focal length 75 mm is placed at 2f distance from the crystal to transfer the signal and idler photon profile to the detectors.

\subsection{Experimentally defining generalized $\sigma_x$ and generalized $\sigma_z$ bases}

\subsection*{Experimental realization of $\sigma_z$}

The state of the down-converted photon after passing through a three-slit can be written as, 
\begin{equation} 
\ket{\psi} = \dfrac{1}{\sqrt{3}} ( c_0 \ket{0} + c_1 \ket{1} + c_2 \ket{2}) \label{eq1}
\end{equation}
The probability to detect a photon prepared in the state $\ket{\psi}$ in the position  corresponding to the $n$-th
slit image is proportional to $|{c_n}|^2$ 
and the measurement operators can be defined as  
\begin{equation}
M_{nf}(n)= \mu_{nf} \ket{n}\bra{n} \label{eq2}
\end{equation}
where $\mu_{nf} $ is the normalization factor\cite{PhysRevA.86.012321}.
The $\sigma_z$ matrix in 3-dimension is 
\begin{equation}
\sigma_z = \begin{bmatrix} 1 & 0 & 0\\ 0 & 0 & 0 \\ 0 & 0 & -1  	\end{bmatrix}
\end{equation}
which can be written as
\begin{equation}
\sigma_z = \ket{0}\bra{0} - \ket{2}\bra{2} \label{eq3} 
\end{equation}
So the positions corresponding to the eigen bases are the center of the first and third slit image profile.

\subsection*{Experimental realization of $ \sigma_x$}

A detection in the position $x$ in the focal plane corresponds to the projector onto $\ket{k_x}$. 
\begin{equation}
k_x =   xk /f \label{eq4}
\end{equation} 
where $k_x$ is the transverse wave vector, $k$ is the wave number and $f$ is the focal length of the lens. The detection probability is proportional to \cite{PhysRevA.86.012321}
\begin{equation}
|\sqrt{\dfrac{a}{2\pi}} \sinc(k_x a/2) \ket{\phi(k_x d)}\bra{\psi}|^2 \label{eq5}
\end{equation} 
where $ \ket{\phi(\theta)} = \ket{0} + \exp{i \theta}\ket{1}  + \exp{2i \theta}\ket{2}$. Hence, the measurement operators in the far field can be defined as $M_{ff}(\theta) \, = \mu_{ff}\, \ket{\phi(\theta)}\bra{\phi(\theta)}$ and the phase parameter is 

\begin{equation}
\theta = 2 \pi x d /\lambda f \label{eq6}
\end{equation}

the operator $ \sigma_x$ can be written as

\begin{equation}
O_1 = \ket{b_0}\bra{b_0} -\ket{b_2}\bra{b_2} \label{eq7}
\end{equation}
where $\ket{b_0}$,$\ket{b_1}$ and $\ket{b_2}$ are 
\begin{equation}
\ket{b_0} =  \dfrac{1}{\sqrt{3}} (\ket{0} + \ket{1} + \ket{2}) \label{eq8}
\end{equation}
\begin{equation}
\ket{b_1} = \dfrac{1}{\sqrt{3}} (\ket{0} + e^{2i\pi/3}\ket{1} + e^{-2i\pi/3}\ket{2}) \label{eq9}
\end{equation}
\begin{equation}
\ket{b_2}= \dfrac{1}{\sqrt{3}} (\ket{0} + e^{-2i\pi/3}\ket{1} + e^{2i\pi/3}\ket{2}) \label{eq10}
\end{equation}
which are also given by the Eqs. (4),(5) and (6) in main paper.

The corresponding angles of the eigenvectors of $\sigma_x$ are ${0, \dfrac{2 \pi}{3}  and \dfrac{4 \pi}{3}}$ respectively. The corresponding detector positions are $0$, $202.5$ and $405$ $\mu m$ which are $x_1$,  $x_2$ and  $x_3$ as mentioned in main paper; where, $\lambda = 0.810 \mu m, f = 7.5 cm, d = 100 \mu m$.



\end{document}